# PRECIPITATION EFFICIENCY AMPLIFIES CLIMATE SENSITIVITY BY ENHANCING TROPICAL CIRCULATION SLOWDOWN AND EASTERN PACIFIC WARMING PATTERN

Ryan L. Li[1]*, Joshua H.P. Studholme[1]*, Alexey V. Fedorov[1,2], Trude Storelvmo[3]

1. Yale University, 2. Sorbonne University, 3. University of Oslo

*Ryan Li and Joshua Studholme share lead author status.

**Email:** ryan.li@yale.edu

**Abstract**

Cloud processes are the largest source of uncertainty in quantifying the global temperature response to $CO_2$ rise. Still, the role of precipitation efficiency (PE) – surface rain per unit column-integrated condensation – is yet to be quantified. Here we use 36 limited-domain cloud resolving simulations from the Radiative-Convective Equilibrium Model Intercomparison Project to show that they strongly imply climate warming will result in increases to net precipitation efficiency. We then analyze 35 General Circulation Models (GCMs) from the Coupled Model Intercomparison Project Phase 6 and find that increasing PE enhances tropical circulation slowdown and strengthens eastern equatorial Pacific warming. These changes trigger pan-tropical positive cloud feedback by causing stratiform anvil cloud reduction and stratocumulus suppression, and thereby amplify overall climate sensitivity. Quantitatively, we find that in the 24 of 35 GCMs which match the cloud-resolving models in simulating increasing PE with greenhouse warming, mean Effective Climate Sensitivity is 1 K higher than in GCMs in which PE decreases. The models simulating increasing PE also comprise all estimates of effective climate sensitivity over 4 K. Taken together, these results show that further constraining PE sensitivity to warming will reduce uncertainty over future climate change.

**Main Text**

**Introduction**

The atmospheric circulation over the tropical Pacific couples intense turbulent ascent in the west to slow stable descent in the east. Over the western tropical Pacific, horizontally extensive ice clouds, known as anvil clouds, cap deep tropical cumuliform clouds and cover a much larger fraction of the troposphere than their lower-level cloud cores. These optically thick cores are Earth's most reflective clouds and thereby represent a significant negative contributor to the planetary energy budget. Above them, the cold anvil clouds have a correspondingly strong positive contribution to the energy budget by radiative absorption (*1*). In the east, shallow stratocumulus clouds mark the vertical boundary between the warm descending air of remote origin and shallow local overturning cells (*2*). These clouds are highly reflective to incoming shortwave radiation, but due to their low altitude and low cloud top temperatures relative to the surface, they have negligible effects on outgoing longwave radiation. The mechanisms governing these two contrasting cloud types are the dominant sources of uncertainty in assessments of Earth's climate sensitivity (*3, 4*), a "multi-trillion-dollar" research question (*5*).

The notion of precipitation efficiency (PE) represents the fraction of condensate to fall to the surface as precipitation, the residual condensate being left behind in the form of clouds. Structurally PE is related to the complex microphysical processes that govern the growth of ice crystals and cloud droplets (6) which can reach the surface as precipitation particles. Thermodynamically, PE represents the fractional latent heat release of atmospheric convection. The other fraction of the total condensate re-evaporates thereby reabsorbing the latent heat released in the original condensation. This links PE to the net energetics of deep convection and thereby to convective updraft strength (7). The mechanisms which set PE largely depend on the humidity of the atmosphere in regions of precipitation. This follows from droplets falling through humid air evaporating less than when falling through drier air (13). However, PE also depends on droplet size, local air temperature and other factors. For example, although tropical mean relative humidity is expected to be constant under greenhouse warming, recent cloud-resolving model (CRM) studies suggest higher PE in warmer climates due to increasing cloud density (14) and convective organization (15, 16).

While other factors influencing atmospheric convection at a range of scales have been studied extensively, such as sea surface temperatures (SSTs, (8)) and free tropospheric humidity (9, 10), the role of PE is poorly understood. This is partly the result of PE being extremely difficult to measure because of the microphysical nature of rain formation, cloud condensation, and evaporation (11). It is also the result of definitions and representations of PE varying immensely across observational and modelling studies (12), making them difficult to synthesize. As a consequence, the role of PE in setting Earth's climate sensitivity is yet to be firmly established.

Previous work (17–20) has proposed somewhat conflicting mechanisms through which PE might impact Effective Climate Sensitivity (ECS) but evidence to support these hypotheses is inconsistent across different global climate models (GCMs; see (11) for a review). Specifically, two studies that imposed an identical increase in PE sensitivity to temperature in different GCMs led to opposing changes in climate sensitivity (18, 19). While simulated cloud responses from imposed

PE changes are structurally complex and model dependent, reduction in cloud liquid and ice water path (i.e., the vertically integrated liquid and ice within an atmospheric column) at higher PE is unequivocal (*18–20*).

In the present study we quantify PE using a parameter $\epsilon$ defined as $\epsilon = P_s/CWP$, where $P_s$ is surface precipitation and $CWP$ is condensed water path (*Methods*; 21). PE is often defined as a fraction of unity by normalizing $P_s$ to a measure of condensation rate, such as the ratio of $P_s$ to condensation in the atmospheric column. However, local condensation is difficult to measure and compute, which leads to a wide range of PE estimates from 0.1 to greater than 1 (*12*). Vigorous hydrological cycling of the atmosphere requires that condensation constantly replenishes the relatively small stock of CWP, and so variations of condensation rate and CWP are closely related. Thus, the parameter $\epsilon$, which is tightly correlated with microphysical measures of PE, captures microphysical cloud condensation at the macrophysical scale and is indeed a measure of PE (21). Unlike other existing PE metrics, $\epsilon$ enables comparable estimates of PE across observations, CRMs and GCMs. For the rest of the manuscript, we use $\epsilon$ and PE interchangeably, and compute climatological $\epsilon$ to represent the net effect of different cloud regimes, from deep convective clouds to non-precipitating shallow clouds, within the tropical cloud ensemble.

The goal of the study is to quantify and explain the effect of PE on the planet's equilibrium temperature response after greenhouse gas increases. Investigating cloud-resolving simulations in the Radiative-Convective Equilibrium Model Intercomparison Project (RCEMIP, (*22*)), we find that they imply PE should increase with warming. To understand what this means for climate sensitivity, we then analyze $\epsilon$ in 35 GCMs participating in Phase 6 of the Coupled Model Intercomparison Project (CMIP6, (*23*)). Using these GCMs we identify two key components linking increasing PE with the resultant warming: (*i*) high cloud feedback associated with retreat and thinning of convective anvils over the western tropical Pacific and (*ii*) low cloud feedback resulting from suppression of stratocumulus decks in the eastern tropical Pacific. These mechanisms are linked via the magnitude of Pacific Walker circulation slowdown, whereby higher PE with greenhouse warming amplifies circulation weakening (*9*). The positive feedback mechanisms (*i*) and (*ii*) act to amplify the overall warming leading to a higher climate sensitivity, while a stronger eastern equatorial Pacific warming pattern mirrors the greater slowdown of the Walker circulation.

**Results**

We begin by considering PE's relationship with surface temperature in 36 models participating in the RCEMIP. These experiments are separated into two domain sizes: 'small' defined as 100 x 100 km$^2$ and 'large' defined as 400 x 6000 km$^2$. Convective self-aggregation is for the most part absent in the small domain simulations but is present in the large domains. The great majority of RCEMIP models predict increasing PE with warming (Fig. 1): 32 out of 36 models in the small domain experiments and 26 out of 28 models in the large domain experiments. The median estimates for $\epsilon$ sensitivity, $4.5\ \%K^{-1}$ in small domains and $3.3\ \%K^{-1}$ in large domains, agree with estimates from positive $\epsilon$ sensitivity GCMs under gradual greenhouse forcing which range from $1$ to $5\ \%K^{-1}$ (*21*). These findings bolster previous single model results that indicated PE will increase with warming (*14*, *15*).

Since cloud-resolving models imply PE increases with warming, we then investigate the significance of this change for large-scale climate. To do this we separate CMIP6 GCMs into two groups based on the sign of the models' tropical-mean $\epsilon$ change following increased atmospheric $CO_2$ concentrations (*Methods*). These $\epsilon$ changes are subsequently referred to as $\epsilon$ sensitivity. Applying this separation criterion cleanly divides CMIP6 models into two sets (Fig. 2A). Dividing these models on either the precipitation sensitivity or CWP sensitivity alone does not yield two distinct groups of models. It has previously been shown that this clean PE separation can be traced to the GCMs' choice of convection parameterization (*21*).

Spatial differences in $\epsilon$ changes between the two model groups are prominent throughout the tropics and in heavy precipitation regions (Fig. 2, B and C). The sign of models' overall $\epsilon$ sensitivity matches up with the sign of the $\epsilon$ change in the Indo-Pacific warm pool (WP; defined as the overlain box in Fig. 2B between 20ºS–20ºN and 80ºE–180ºE). In the equatorial Eastern Pacific (EP; 30ºS–30ºN and 80ºW–180ºW), $\epsilon$ increases in both model groups but in models with positive $\epsilon$ sensitivity this $\epsilon$ increase is twofold greater than in models with negative $\epsilon$ sensitivity.

We find that $\epsilon$ sensitivity is proportional to the local net cloud feedback in both the WP and EP ($r = 0.60$ and $0.51$ respectively; Fig. 2D). In the WP, the cloud feedback change is dominated by the shortwave component ($1.44\ W\ m^{-2}\ K^{-1}$) but it is partially compensated by the longwave component ($-0.69\ W\ m^{-2}\ K^{-1}$), summing to a net feedback of $0.75\ W\ m^{-2}\ K^{-1}$ (Fig. S1). The dominance of the shortwave feedback means that the sign of the overall WP cloud feedback opposes the sign expected by the proposed infrared iris effect (18). This agrees with results of other GCM sensitivity studies that imposed PE increases (*19, 20*). This PE–cloud feedback relationship suggests a mechanism whereby enhanced convective rainout limits detrainment and the reduced anvil cloud source results in their contraction and thinning.

The similar correlation magnitudes between $\epsilon$ sensitivity and net cloud feedback in both the WP and EP hint at a mechanistic link coupling the cloud feedbacks across the entire Pacific (Fig. S4). The net cloud feedback difference over the EP between GCMs with positive vs negative $\epsilon$ sensitivity is also dominated by the shortwave component ($1.18\ W\ m^{-2}\ K^{-1}$, Fig. S1). This is also slightly offset by a longwave component ($-0.36\ W\ m^{-2}\ K^{-1}$), and sums to a net value of $0.82\ W\ m^{-2}\ K^{-1}$.

Consequently, CMIP6 models with positive $\epsilon$ sensitivity exhibit stronger positive cloud feedbacks in both the WP and EP relative to their counterparts with negative $\epsilon$ sensitivity (Fig. S1). In the average over the entire tropics, the net cloud feedback difference between the two model groups is $0.49\ W\ m^{-2}\ K^{-1}$ (Fig. S1A). Again, this follows from larger positive shortwave feedback ($0.89\ W\ m^{-2}\ K^{-1}$, Fig. S1B) being partly compensated by smaller negative longwave feedback ($-0.40\ W\ m^{-2}\ K^{-1}$, Fig. S1C). A decomposition analysis by cloud height (*Methods*) reveals the dominant contributions come from medium and high clouds in the WP and low clouds in the EP (Fig. S3). The WP medium and high net cloud feedback of $0.40\ W\ m^{-2}\ K^{-1}$ (Fig. S3A) and the EP low cloud net feedback of $0.51\ W\ m^{-2}\ K^{-1}$ (Fig. S3D) are tightly linked (correlation coefficient $r = 0.64$, Fig. S4).

We connect this array of cloud differences across the Pacific to the concomitant enhanced slowdown of the Walker circulation in positive vs. negative $\epsilon$ sensitivity GCMs (Fig. 3). The differences in the large-scale circulation slowdown are statistically significant in the ascending WP and subsiding EP (Fig. 3F). In the convectively active WP, anvil cloud fraction reduces as the planet warms and the large-scale circulation slows (Fig. 3, A and B), a feature of both groups of GCM simulations. The reduction of cloud cover peaks near the climatological anvil cloud maximum at 120 ºE and 200 hPa, where there is a strong difference in the magnitude of this effect between positive (15%) and negative (5%) $\epsilon$ sensitivity models (Fig. 3C). From the top of the atmosphere, these simulated changes appear as a contraction in the area covered by anvil clouds as well as thinning (Fig. S5) since the clouds are less opaque to sunlight. In the EP, the radiatively bright stratocumulus clouds are capped from above by the descending branch of the Walker circulation. In our results, associated with the enhanced large-scale circulation slowdown in positive $\epsilon$ sensitivity models, these EP stratocumulus decks are suppressed (Fig. 3C). This triggers a very strong local shortwave cloud feedback (Fig. S3C). We think this is very likely the dominant cause of the enhanced southern hemisphere Eastern Pacific SST warming (Fig. 3I).

With enhanced Walker circulation slowdown by increasing PE, the equatorial surface easterlies are correspondingly weakened (Fig. S6). It thus follows that this reduced wind stress dampens the climatological oceanic upwelling at the equator and off South America's western coast. This second order effect of increasing PE is evident in an El Nino-like SST warming pattern (Fig. 3I). This adds to the Eastern Pacific warming caused by the local stratocumulus suppression. In addition to these two effects, warmer SSTs and weaker surface temperature inversion (Fig. S7) create less favorable conditions for stratocumulus decks (*2*). Amplification of these effects in positive $\epsilon$ sensitivity GCMs further contributes to the reduction of low clouds and thus establishes the positive cloud feedback in the EP (Fig. S3E).

Given that $\epsilon$ is specifically associated with deep convection, one assumption might be that the WP, where climatological deep convection is strongest, is in the driving seat of PE's influence. However, WP anvil clouds have cold cloud tops, and reduced fractional cover accompany local negative longwave feedback (Fig. S3C). This partly compensates the positive shortwave feedback (Fig. S3B) and thus dampens the net anvil cloud feedback (Fig. S3A). On the other hand, climatological PE changes are also strong in positive $\epsilon$ sensitivity models in the low cloud dominant EP (Fig. 2B). This implies there could be changes in transient deep convection in the EP, a climatological region of descent. Together, we interpret PE's influence as being pan-tropical and profoundly coupled to the clouds and large-scale convective circulation.

The Walker circulation connects a diverse array of tropical cloud regimes. In the WP, positive anvil cloud feedback amplifies the local warming, which is enhanced at higher altitudes by moist adiabatic adjustment (Fig. S8). The resulting weaker temperature stratification is communicated throughout the tropics as the weak effective planetary rotation at low latitudes cannot sustain horizontal temperature gradients (*24*). With higher PE, measured by increased $\epsilon$, updrafts become more efficient, wherein a weaker circulation sustains the same latent heating (Fig. 3, D–F). The Walker circulation bridges the WP and EP cloud feedback, and, under greenhouse warming, increasing $\epsilon$ causes greater Walker circulation slowdown, amplifying the positive cloud feedback.

Collectively these results reveal that the integrated links between enhanced large-scale circulation slowdown and the cloud changes across the Pacific associated with PE changes result in an overall positive cloud feedback. This suggests that PE amplifies warming following increases in greenhouse gases. We quantify this effect by considering these models ECS values and find that positive $\epsilon$ sensitivity is a necessary but insufficient condition for high ECS (Fig. 4A). In the 24 out of the total 35 CMIP6 GCMs which match the cloud-resolving models in simulating increasing PE with greenhouse warming, mean ECS is 1 K higher than in GCMs in which PE decreases. While models with positive $\epsilon$ sensitivity exhibit both high and low ECS estimates, all negative $\epsilon$ sensitivity models have low end ECS. Moreover, all GCMs with ECS greater than 4 K (10 of 35 models) have positive $\epsilon$ sensitivity (Fig. 4A).

We decompose ECS into its individual feedback components to confirm that the above mentioned ECS differences are dominated by tropical clouds. Among all possible feedbacks, only assuming zero cloud feedback results in the two model groups becoming statistically indistinguishable (Fig. 4B). Furthermore, the two model groups become statistically indistinguishable in a separate ECS calculation assuming zero tropical cloud feedback (Fig. 4A). When tropical clouds are excluded from the ECS calculation, all ECS estimates above 4 K disappear, affirming the critical importance of tropical cloud feedback for very high climate sensitivity.

**Discussion**

We have shown that the vast majority of CRMs suggest that the PE of deep convective clouds will increase under anthropogenic climate change. These multi-model RCEMIP results agree with recent single model CRM studies (14, 15). Such studies have found that at higher surface temperatures, denser clouds support higher fractional rainout (14). Re-evaporation of rain is also reduced at warmer temperatures (25). Moreover, GCMs predict that changes in the large-scale tropical circulation pattern with warming will increase rainfall unevenness (26), i.e., there is more convective organization. Further, satellite records, albeit short, show tentative signs that these anticipated changes in $\epsilon$ are already appearing (Fig. S9). This may be the result of increases in the frequency of organized deep convection in the 21st century as recently identified in satellite observations (*27, 28*).

The amplified Walker circulation slowdown in models with increasing PE is coupled to the development of a stronger eastern equatorial Pacific warming pattern (Fig. 3; I, G, and H). This pronounced El Niño-like mean pattern, extending from the eastern Pacific to the dateline along the equator (29, 30), is projected to emerge with future warming but its strength varies greatly across the models (31). Thus, the PE considerations provide a potential constraint on the strength of this pattern in the future.

Projecting the radiative effects of Earth's deep convective anvils in different climates has been a major challenge. It remains unclear why the net radiative effect of anvils is near neutral in the present day because these clouds are strongly influenced by processes that remain unresolved in contemporary models (*32*). Nevertheless, in a warmer climate, high clouds remain near the same temperature and thus rise in altitude (*33*). As the clouds rise, temperature stratification increases, and this may result in less mass divergence from convective updrafts. This thermodynamic effect has been hypothesized to reduce tropical high cloud cover (*34*). While we have verified that this effect is indeed present in the CMIP6 models, we find that it is independent from the PE–climate sensitivity mechanism identified in this work (Fig. S10).

As convective clouds become more efficient at precipitating in a warmer atmosphere, the Walker circulation slows, and powerful positive cloud feedbacks proportional to the PE increase emerge in the western and eastern Pacific. In the Indo-Pacific warm pool, there is contraction and thinning of high clouds. Simultaneously, reduced subsidence triggers stratocumulus suppression which warms underlying SSTs, further weakening stratocumulus decks. The dampened equatorial wind stress associated with enhanced Walker circulation slowdown reduces oceanic upwelling in the EP, which further adds to SST increases and thus the local stratocumulus suppression. Coupling between the Walker circulation slowdown and SST enhances the eastern equatorial

Pacific warming pattern. Together, the coupled tropical Pacific cloud feedbacks are positive, causing high climate sensitivity (ECS > 4 K) in state-of-the-art GCMs. Thus, we have shown that changes in the precipitation efficiency of deep convective clouds controls tropical cloud feedback via the rates of tropical circulation slowdown and are critical for constraining spatial patterns of climate change and, in aggregate, climate sensitivity.

**Materials and Methods**

Precipitation Efficiency measure

We use the measure of precipitation efficiency (PE) $\epsilon$ (21):

$$\epsilon = \frac{P_s}{CWP},$$

where $P_s$ is the surface precipitation rate and $CWP$ is vertically integrated condensed water and ice in the atmospheric column. $\epsilon$ has high correspondence with the microphysical definition of PE. Its inverse $\epsilon^{-1} = \tau_c$ measures the average residence time scale of clouds in the atmosphere, where high $\epsilon$ is equivalent to low $\tau_c$ signaling shorter residence time of clouds and high precipitation efficiency.

While models with positive $\epsilon$ sensitivity generally show decreases in $CWP$ and higher fractional increases in $P_s$, the difference in $\epsilon$ sensitivity cannot be explained by $P_s$ nor $CWP$ alone. Rather, differences in the greenhouse warming response of $\epsilon$ encapsulates a wide range of $P_s$ and $CWP$ perturbations (Fig. 1A and Fig. S2). The large $CWP$ range reflects differences in cloud parameterizations that have long been recognized as the dominant source of uncertainty in future climate projections (4). Tropical $\epsilon$ sensitivity is proportional to tropical-mean $\epsilon$ (Fig. S11). CMIP6 models show a wide tropical-mean $\epsilon$ range in the present climate. Understanding controls on tropical-mean $\epsilon$ and hence the magnitude of $\epsilon$ sensitivity to temperature deserves future research.

CMIP6 data

Thirty-five models from the Coupled Model Intercomparison Project – Phase 6 (CMIP6) are used in this study to compute $\epsilon$, all models with available output in the CMIP6 data repository in both the preindustrial control (piControl) and abrupt quadrupling of atmospheric $CO_2$ (abrupt-4x$CO_2$) scenarios. Sensitivity of $\epsilon$ to greenhouse warming is defined as the change in $\epsilon$ spatially averaged from 30 ºS to 30 ºN and temporally averaged over the final 50 years in abrupt-4x$CO_2$ relative to piControl.

*Models with negative $\epsilon$ sensitivity (total 11):*

AWI-CM-1-1-MR, CAMS-CSM1-0, FGOALS-g3, GFDL-CM4, GFDL-ESM4, GISS-E2-1-G, GISS-E2-1-H, GISS-E2-2-G, MPI-ESM-1-2-HAM, MPI-ESM1-2-HR, and MPI-ESM1-2-LR.

*Models with positive $\epsilon$ sensitivity (total 24):*

BCC-CSM2-MR, BCC-ESM1, CESM2, CESM2-FV2, CESM2-WACCM, CESM2-WACCM-FV2, CIESM, CMCC-CM2-SR5, CMCC-ESM2, CanESM5, E3SM-1-0, EC-Earth3-AerChem, IITM-ESM, INM-CM4-8, INM-CM5-0, IPSL-CM5A2-INCA, IPSL-CM6A-LR, KACE-1-0-G, MIROC6, MRI-ESM2-0, NorESM2-LM, NorESM2-MM, SAM0-UNICON, and TaiESM.

Environmental metrics

For specific environmental variables used in the analysis, all 35 models with $\epsilon$ data are not always available. The number of models with available output are provided in brackets in the title of each subplot, these models are always subsets of the lists above and statistical significance is always assessed against the same robustness level. Profiles of cloud fraction (Fig. 3) and cloud liquid and ice content (Fig. S6) are interpolated from each model's native pressure grid to a designated vertical grid with 47 levels. GISS models are not included in the multi-model mean cloud liquid and ice content profiles due to magnitude errors such as non-physical values and missing data.

Effective Climate Sensitivity (ECS) estimates, defined as the negative ratio of the effective radiative forcing from doubling of carbon dioxide to the net climate feedback parameter, are obtained from (*35*) which are diagnosed based on the offline radiative kernels of (*36*). These data are publicly available at: https://github.com/mzelinka/cmip56_forcing_feedback_ecs. Cloud feedbacks (Fig. 2D, Fig. S3, and Fig. S4) are diagnosed using offline radiative kernels and provided by Dr. Mark Zelinka, who used the approximate partial radiative perturbation technique to separate the shortwave cloud feedback into its amount and scattering components and further decomposed the cloud feedbacks into contributions from low and non-low clouds (*35*).

The Estimated Inversion Strength (EIS) measures the strength of the boundary layer inversion, a critical environmental control for low cloud formation (*2*). EIS is defined as:

$$EIS = \theta_{700} - \theta_0 - \Gamma_m^{850}(z_{700} - LCL),$$

where $\theta_{700}$ and $\theta_0$ are potential temperature at 700 hPa and at the surface, respectively, $\Gamma_m^{850}$ is the moist adiabatic lapse rate at 850 hPa, $z_{700}$ is the height of the 700 hPa surface, and $LCL$ is the lifting condensation level. High EIS values correspond to favorable conditions for stratocumulus decks. Outside the eastern tropical and subtropical Pacific, there are very few stratocumulus and shallow cumulus clouds (*37*).

The stability-iris hypothesis (*34*) suggests that the radiatively-driven mass divergence in clear-sky regions $D_r$ is a determining factor for anvil cloud fraction. $D_r$ is given by:

$$D_r = \frac{\partial \omega}{\partial P},$$

where $\omega$ is the vertical pressure velocity and $P$ is pressure. $D_r$ when summed over all subsidence regions ($\omega > 0$), it is a measure of convective mass divergence. The response of mass divergence to warming is essentially indistinguishable between positive and negative $\epsilon$ sensitivity GCMs (Fig. S7). Thus, this mechanism cannot explain the differences in the anvil cloud fraction response across positive and negative $\epsilon$ sensitivity models. In positive $\epsilon$ sensitivity CMIP6 GCMs, the amount of detrained condensate associated with a given mass divergence reduces with warming.

Cloud resolving model estimates of $\epsilon$

We use results from the CRM experiments in the Radiative-Convective Equilibrium Model Intercomparison Project (RCEMIP), which are run on a doubly-periodic domain with perpetual sunlight, fixed and uniform SSTs, and no rotation (22). Small domain simulations use a square domain of 100 km by 100 km and a horizontal resolution of 1 km. Large domain simulations employ a channel geometry of 400 km by 6000 km, allowing the possibility of convective organization. Domain-averaged diagnostics of precipitation and condensed water path for small and large domain simulations at 295K, 300K, and 305K SSTs are provided by Dr. Allison Wing. Fractional change in $\epsilon$, $\Delta\epsilon/\epsilon$, for each model is computed relative to the 295K SST simulation.

Satellite observations

Observations of $\epsilon$ trends (Fig. S9) are obtained by combining satellite measurements of $CWP$ and $P_s$. $P_s$ is obtained from the Tropical Rainfall Measuring Mission (TRMM) Multisatellite Precipitation Analysis (TMPA) 3B42 monthly precipitation product (*38*) by the National Aeronautics and Space Administration (NASA) and the Japan Aerospace Exploration Agency (JAXA), re-gridded to 1° by 1° resolution. Two independent $CWP$ observations are used: (*i*) the Advanced Very High Resolution Radiometer post meridiem (AVHRR-PM) dataset version 3 (*39*) as part of the European Space Agency (ESA) Climate Change Initiative (CCI) project and (*ii*) the MODerate-resolution Imaging Spectroradiometer (MODIS) Level 3 Atmosphere Monthly Product (*40*). Both $CWP$ datasets are at 1° by 1° horizontal resolution. $CWP$ is estimated in both satellite products using the following expressions (*41*):

$$IWP = \frac{4}{3}\frac{\tau_i R_{ei} \rho_i}{Q_i},$$

$$LWP = \frac{4}{3}\frac{\tau_l R_{el} \rho_l}{Q_l},$$

where $\tau$ is optical thickness, $R_e$ is the effective radius, $\rho$ is density, and Q is the extinction coefficient, evaluated for liquid (subscript *l*) and ice clouds (subscript *i*) separately. $CWP$ is the sum of $IWP$ and $LWP$, then converted to values comparable to GCM grid-box averages by multiplying by the observed cloud fraction.

**Acknowledgments**

The authors thank Dr. Mark Zelinka (Lawrence Livermore National Laboratory) for graciously providing us the decomposed cloud feedbacks and making CMIP6 ECS data publicly available. We also thank Dr. Allison Wing for kindly providing us the domain-averaged diagnostics of precipitation and condensed water path from the RCEMIP archive. We thank Dr. Nicholas Lutsko (Scripps Institution for Oceanography), Dr. Shineng Hu (Duke University), and Dr. Ivy Tan (McGill University) for helpful discussions of this project.

**Figures and Tables**

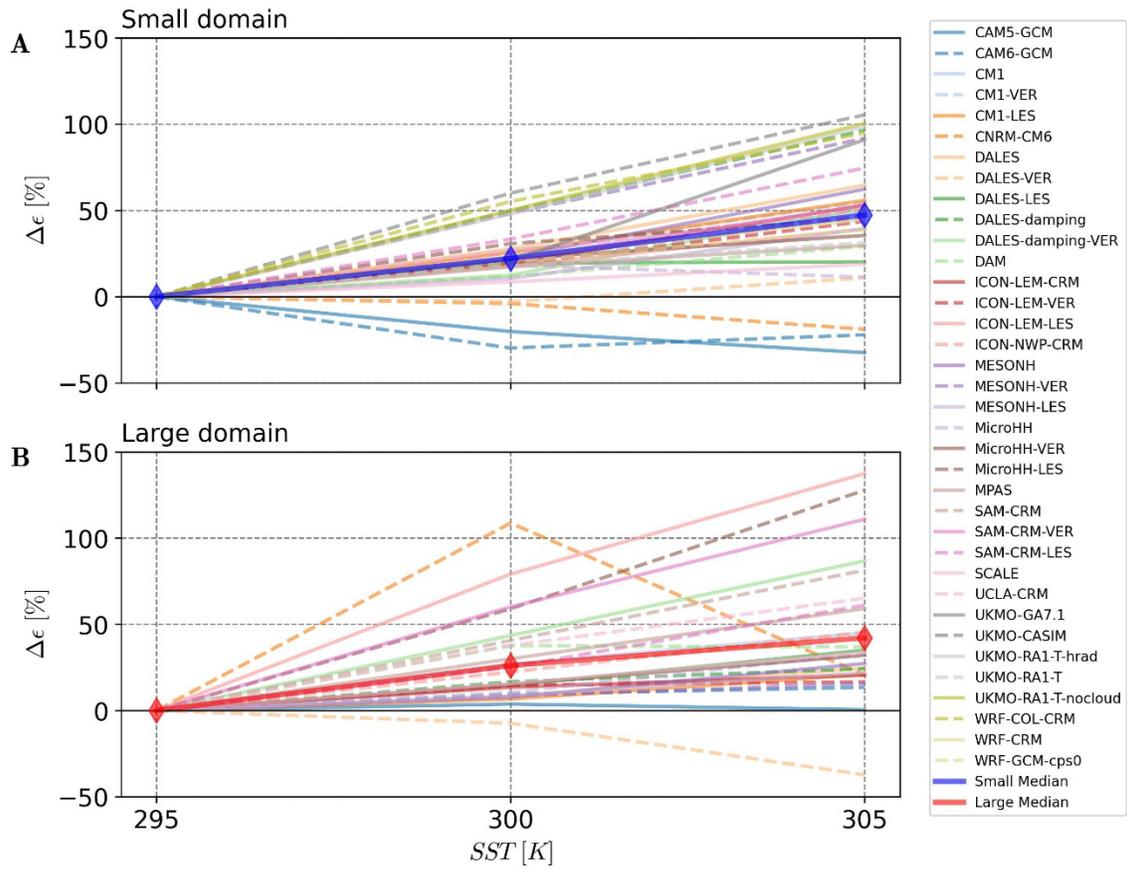

**Figure 1. Cloud-resolving model ensemble indicate positive ϵ sensitivity.** Change in precipitation efficiency Δϵ simulated in (A) small and (B) large domains by large-eddy simulations, cloud-resolving models, single-column models, and general circulation models participating in the RCEMIP. Among the 36 small and 28 large domain experiments available, 32 small domain models and 26 large domain models show increasing PE with warming. Change is relative to the 295K simulation. Diamonds show median ϵ sensitivity across all models.

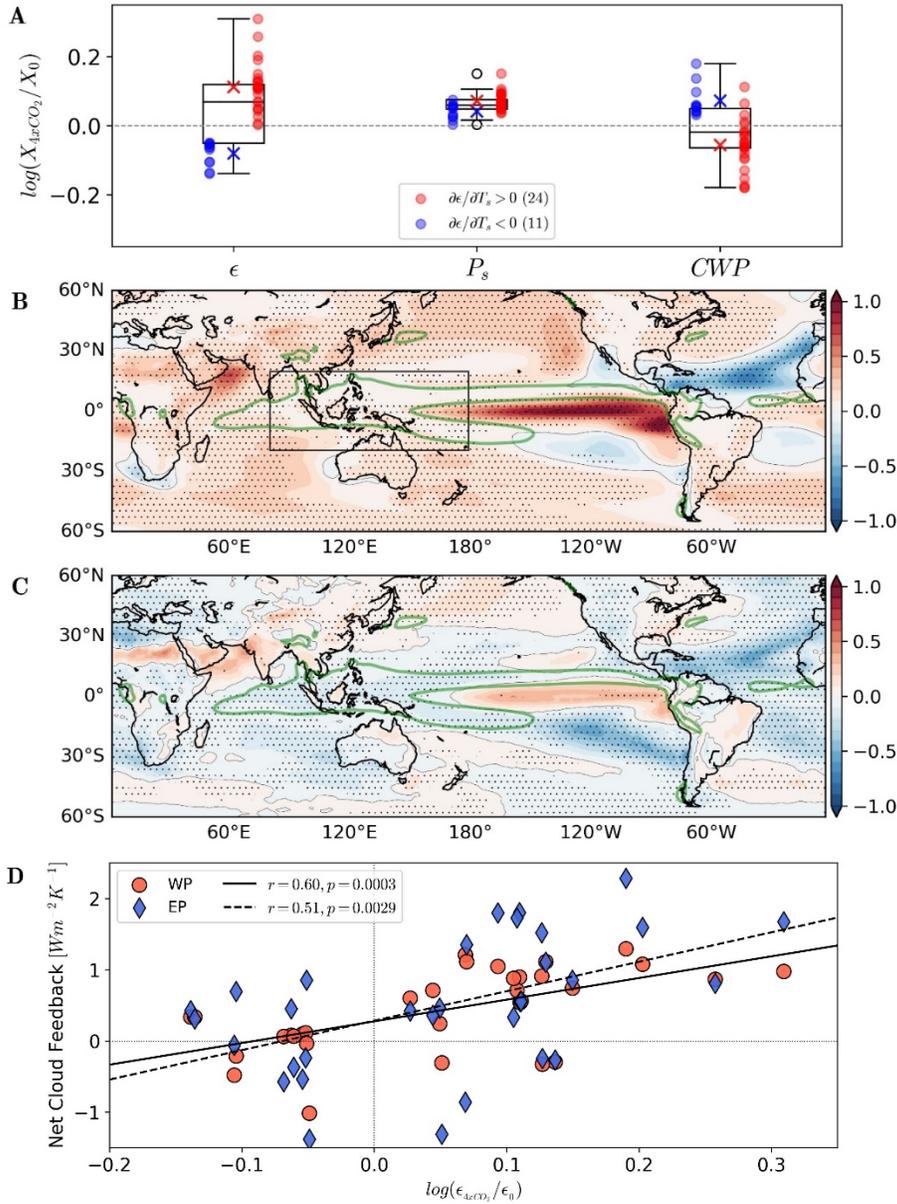

**Figure 2. Impact of the disagreement of precipitation efficiency sensitivity in CMIP6.** (A) Sensitivity of tropical $\epsilon$, precipitation ($P_s$), and condensed water path ($CWP$) to greenhouse warming for two groups of climate models: with positive (red) and with negative (blue) sensitivities of tropical-mean (30˚S–30˚N) $\epsilon$ to greenhouse warming. (B, C) Spatial maps of multi-model mean $\epsilon$ sensitivity for models with positive and with negative $\epsilon$ sensitivities, respectively. Grey shading in (A) indicates lack of statistical significance (p-value ≥ 0.05). Stippling in (B) and (C) indicates more than 75% of models agree on the sign of the response. (D) Net cloud feedback estimated over the Indo-Pacific Warm Pool (WP; circles) and over the Eastern Pacific (EP; diamonds) scattered against tropical $\epsilon$ sensitivity to greenhouse warming for different models. The solid and dashed lines represent linear regressions for WP and EP data, respectively, where r is the Pearson correlation coefficient and p is the associated *p* value. Overall, higher values of cloud feedback correspond to positive $\epsilon$ sensitivity. See *Methods* for CMIP6 model and simulation details.

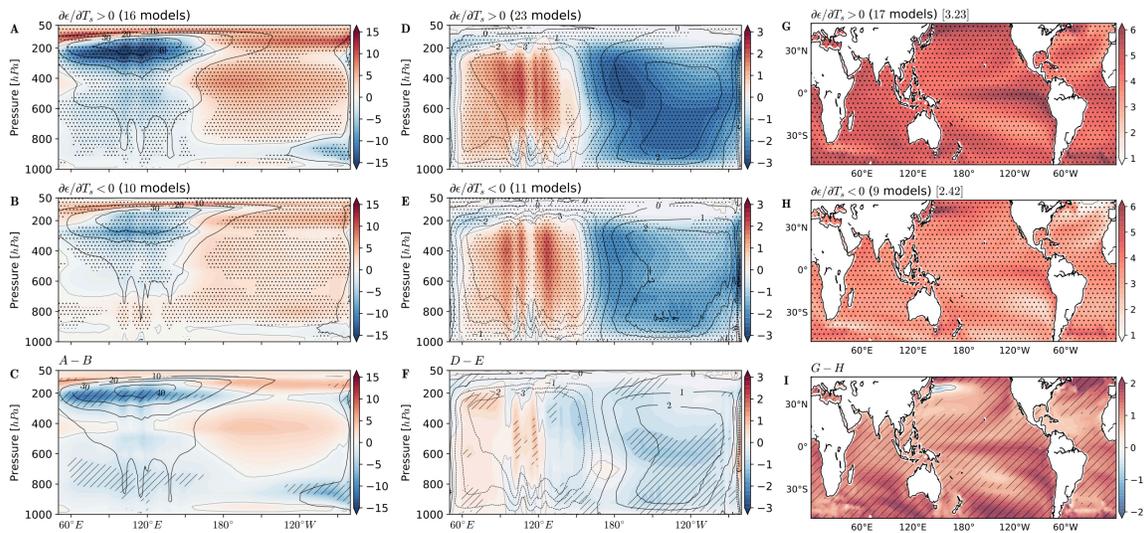

**Figure 3. Contraction and thinning of deep convective anvil clouds, low stratocumulus cloud suppression, Walker circulation weakening, and enhanced Eastern Pacific warming.** Changes in cloud areal coverage (cloud fraction) along the equator in response to greenhouse warming in (A) positive and (B) negative ϵ sensitivity models, and (C) the difference between the two, i.e., panel (A) minus panel (B). (D–F) As the left column but for changes in vertical pressure velocity. (G–I) As the left column but for changes in sea surface temperature. Black contours in (A–F) show multi-model mean preindustrial climatology. In the titles of panels, square brackets indicate the tropical mean SST (30ºS–30ºN) and round brackets indicate the number of models. Stippling indicates more than 75% of models agree on the sign of the response; hatching shows p-value less than 0.05 using Welch's t-test. Cloud fraction is given in % spatial coverage, velocity in $10^{-2}$ pascals per second, and SST in K. Experiments with abrupt quadrupling of atmospheric $CO_2$ relative to preindustrial control have been used. For the left and middle columns, the data is averaged between 5ºS – 5ºN.

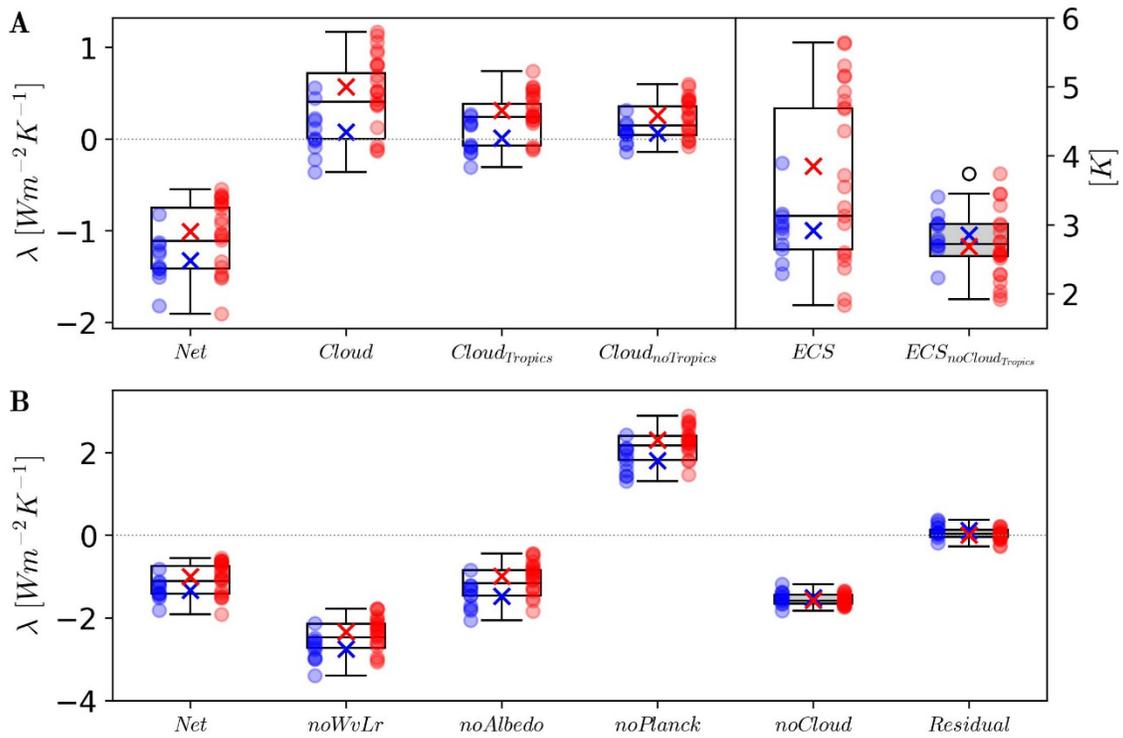

**Fig. 4. Role of tropical cloud feedback in ECS spread.**

(A) The Net feedback (sum of all individual climate feedbacks) and Cloud feedback for 35 GCMs employed in this study. The Cloud feedback is decomposed into contributions from tropical clouds (30ºS–30ºN; $Cloud_{Tropics}$) and extratropical clouds (90–30ºS and 30–90ºN; $Cloud_{Extratropics}$). Effective Climate Sensitivity ($ECS$) is defined as the effective radiative forcing of CO2 doubling divided by the net climate feedback parameter (35). The multi-model mean ECS for positive and negative $\epsilon$ sensitivity models are $3.84\ K$ and $2.91\ K$, respectively. A hypothetical ECS estimate that assumes zero tropical cloud feedback ($ECS_{NoCloud_{Tropics}}$) is also provided. (B) The Net feedback with each individual feedback component removed: water vapor and lapse rate (noWvLr), albedo (noAlbedo), Planck (noPlanck), and cloud (noCloud). Without cloud feedback, the feedback parameters are indistinguishable from the residual term in the linear framework. Units are given in square brackets.